%% file: main.tex
\documentclass[trackchanges,twocolumn]{aastex701}

\usepackage{newunicodechar}
\usepackage{amsmath}
\usepackage{color}
\usepackage{booktabs}
\usepackage{rotating}
\usepackage{longtable}  
\DeclareUnicodeCharacter {2212} {-}
\DeclareUnicodeCharacter {02BC} {-}
\submitjournal{ApJ}
\accepted{January 9, 2026}

\begin{document}

\title{Digging into the Interior of Hot Cores with ALMA (DIHCA). VI. The Formation of Low-mass Multiple Systems in High-mass Cluster-forming Regions}


\author[0000-0003-4506-3171]{Qiu-yi Luo}\email{luoqiuyi233@gmail.com}
\affiliation{Institute of Astronomy, Graduate School of Science, The University of Tokyo, 2-21-1 Osawa, Mitaka, Tokyo 181-0015, Japan}
\affiliation{Department of Astronomy, School of Science, The University of Tokyo, 7-3-1 Hongo, Bunkyo, Tokyo 113-0033, Japan}

\author[0000-0002-7125-7685]{Patricio Sanhueza}\email{patosanhueza@gmail.com}
\affiliation{Department of Astronomy, School of Science, The University of Tokyo, 7-3-1 Hongo, Bunkyo, Tokyo 113-0033, Japan}

\author[0000-0003-1252-9916]{Stella S. R. Offner}\email{soffner@astro.as.utexas.edu}
\affiliation{Department of Astronomy, The University of Texas at Austin, Austin, 
152-8551, USA.}

\author[0000-0002-8250-6827
]{Fernando Olguin}\email{f.olguin@yukawa.kyoto-u.ac.jp}
\affiliation{Center for Gravitational Physics, Yukawa Institute for Theoretical Physics, Kyoto University, Kitashirakawa Oiwakecho, Sakyo-ku, Kyoto 606-8502, Japan}
\affiliation{National Astronomical Observatory of Japan, National Institutes of Natural Sciences, 2-21-1 Osawa, Mitaka, Tokyo 181-8588, Japan}

\author[0000-0001-6431-9633]{Adam Ginsburg}\email{adam.g.ginsburg@gmail.com}
\affiliation{Department of Astronomy, University of Florida, P.O. Box 112055, Gainesville, 32611-2055, USA.}

\author[0000-0001-5431-2294]{Fumitaka Nakamura}\email{fumitaka.nakamura@nao.ac.jp}
\affiliation{Department of Astronomy, Graduate School of Science, The University of Tokyo, 7-3-1 Hongo, Tokyo, 113-0033, Japan }
\affiliation{Division of Science, National Astronomical Observatory of Japan, Osawa 2-21-1, Mitaka, 181-8588, Japan}

\author[0000-0002-6752-6061]{Kaho Morii}\email{kaho.morii@cfa.harvard.edu}
\affiliation{Center for Astrophysics $|$ Harvard \& Smithsonian, 60 Garden Street, Cambridge, MA 02138, USA}
\affiliation{Department of Astronomy, Graduate School of Science, The University of Tokyo, 7-3-1 Hongo, Bunkyo-ku, Tokyo 113-0033, Japan}
\affiliation{National Astronomical Observatory of Japan, National Institutes of Natural Sciences, 2-21-1 Osawa, Mitaka, Tokyo 181-8588, Japan}

\author[0000-0002-8691-4588
]{Yu Cheng}\email{yu.cheng@astro.nao.ac.jp}
\affiliation{Division of Science, National Astronomical Observatory of Japan, Osawa 2-21-1, Mitaka, 181-8588, Japan}

\author[0000-0002-6907-0926
]{Kei E. I. Tanaka}\email{kt503i@gmail.com}
\affiliation{Department of Earth and Planetary Sciences, Institute of Science Tokyo, Meguro, Tokyo, 152-8551, Japan}

\author[0000-0002-4774-2998
]{Junhao Liu}\email{junhao.liu@nao.ac.jp}
\affiliation{Division of Science, National Astronomical Observatory of Japan, Osawa 2-21-1, Mitaka, 181-8588, Japan}

\author[0000-0002-5286-2564]{Tie Liu}\email{liutie@shao.ac.cn}
\affiliation{Shanghai Astronomical Observatory, Chinese Academy of Sciences, 80 Nandan Road, Shanghai 200030, People’s Republic of China}

\author[0000-0003-2619-9305]{Xing Lu}\email{xinglu@shao.ac.cn}
\affiliation{Shanghai Astronomical Observatory, Chinese Academy of Sciences, 80 Nandan Road, Shanghai 200030, People’s Republic of China}

\author[0000-0003-2384-6589]{Qizhou Zhang}\email{qzhang@cfa.harvard.edu}
\affiliation{Center for Astrophysics $|$ Harvard \& Smithsonian, 60 Garden Street, Cambridge, MA 02138, USA}

\author[0000-0003-4402-6475]{Kotomi Taniguchi}\email{kotomi.taniguchi@nao.ac.jp}
\affiliation{Division of Science, National Astronomical Observatory of Japan, Osawa 2-21-1, Mitaka, 181-8588, Japan}

\author[0000-0002-0028-1354]{Piyali Saha}\email{s.piyali16@gmail.com}
\affiliation{Academia Sinica Institute of Astronomy and Astrophysics, No.1, Sec. 4., Roosevelt Road, Taipei 10617, Taiwan
}
\affiliation{ National Astronomical Observatory of Japan, National Institutes of Natural Sciences, 2-21-1 Osawa, Mitaka, Tokyo 181-8588, Japan}

\author[0000-0003-1275-5251]{Shanghuo Li}\email{shanghuo.li@gmail.com}
\affiliation{School of Astronomy and Space Science, Nanjing University, 163 Xianlin Avenue, Nanjing 210023, People's Republic of China}
\affiliation{Key Laboratory of Modern Astronomy and Astrophysics (Nanjing University), Ministry of Education, Nanjing 210023, People's Republic of China}

\author[0000-0001-7573-0145]{Xiaofeng Mai}\email{maixf1418@gmail.com}
\affiliation{Shanghai Astronomical Observatory, Chinese Academy of Sciences, 80 Nandan Road, Shanghai 200030, People’s Republic of China}

\begin{abstract}

Most stars form in multiple systems, with profound implications in numerous astronomical phenomena intrinsically linked to multiplicity. However, our knowledge about the process on how multiple stellar systems form is incomplete and biased toward nearby molecular clouds forming only low-mass stars, which are unrepresentative of the stellar population in the Galaxy. 
Most stars form within dense cores in clusters alongside high-mass stars ($>$8 M$_{\odot}$), as likely the Sun did.  Here we report deep ALMA 1.33 mm dust continuum observations at $\sim$160 au spatial resolution, revealing 72 low-mass multiple systems embedded in 23 high-mass cluster-forming regions, as part of the Digging into the Interior of Hot Cores with ALMA (DIHCA) survey. We find that the companion separation distribution presents a distinct peak at $\sim$1200 au, in contrast to the one at $\sim$4000 au observed in nearby low-mass regions. The shorter fragmentation scale can be explained by considering the higher pressure exerted by the surrounding medium, which is higher than the one in low-mass regions, due to the larger turbulence and densities involved. Because the peak of the companion separation distribution occurs at much larger scales than the expected disk sizes, we argue that the observed fragmentation is produced by turbulent core fragmentation. Contrary as predicted, the multiplicity fraction remains constant as the stellar density increases. We propose that in the extremely dense environments where high-mass stars form, dynamical interactions play an important role in disrupting weakly bound systems.

\end{abstract}

\section{Introduction}\label{sec:intro}

Multiple systems play a crucial role in stellar evolution, influencing processes from planet formation and orbital dynamics to supernova explosions and the formation of compact binaries, which are so far the most common sources of observed gravitational wave events \citep{2023Tauris}. 
The formation of multiple systems is generally explained by three pathways: core fragmentation, disk fragmentation, and dynamical capture \citep{Offner23}. Core fragmentation occurs in the early stages, from the prestellar to the protostellar phase, where turbulence triggers multiple collapses individually within a single dense core on scales of thousands of au \citep{Goodwin04,Goodwin07}. Disk fragmentation takes place during the accretion phase, when parts of the accretion disk become gravitationally unstable (i.e., when the Toomre Q parameter $\textless$ 1) \citep{Tohline02,Clark10,Kratter16}. The resulting fragments typically form on scales of $\sim$100 au, consistent with disk sizes. Dynamical capture occurs when an initially unbound protostar becomes gravitationally bound to a nearby stellar system \citep{Ostriker94}. 
Capturing the kinematic information of each protostar in  multiple systems is challenging, so these formation pathways are usually distinguished by their characteristic fragmentation scales \citep{Offner23}.

Among the three, core fragmentation is considered a primary pathway for forming multiple systems \citep{kh23}. 
ALMA observations of fragmentation in starless cores have been conducted in nearby clouds such as Chamaeleon and Ophiuchus \citep{Dunham16,Kirk17}. \cite{Ohashi18} and \citet{Caselli19} identified multiple density peaks within individual  prestellar cores, TUKH122 and L1544, respectively. 
More recently, \citet{Sahu21} confirmed the presence of substructures within a prestellar core with sufficient mass ($\textgreater$ 0.1 M$_{\odot}$) and size ($\sim$ 1200 au) to form protostars in the Orion cloud.
Previous surveys toward the Perseus and Orion clouds have reported multiplicity fractions of $\sim$30\% and 44\% \citep{Tobin16, Tobin22} . These surveys reveal similar bimodal companion separation distributions, with one peak at 75 au, which likely results from disk fragmentation with a contribution of cores that formed via core fragmentation that migrated inward; while another peak at 4000 au is explained by primordial core fragmentation. 
\citet{Luo22} suggests that both thermal Jeans fragmentation and turbulent fragmentation contribute to multiple system formation in star-forming cores in the Orion region predominantly forming low-mass stars. 
However, existing studies have mostly focused on nearby clouds (distance $\textless$ 500 pc) that only form low-mass stars.

High-mass cluster-forming (HMCF) regions ($\sim$1 pc structures forming both low- and high-mass stars) are considered representative sites of star formation in our Galaxy.
Clump fragmentation from pc to a few thousand au in these regions is often explained by thermal Jeans fragmentation \citep{Sanhueza17,Sanhueza19,Beuther18,Morii23,Morii24,Ishihara24}, however, how  cores (a few $\sim$1000 au size objects) fragment into bound binary or higher-order multiple systems is unclear.  
The DIHCA (Digging into the Interior of Hot Cores with ALMA) survey covers 30 high-mass cluster-forming regions. \citet{Li24} studied multiplicity in one examplar region, G333.23--0.06, and \citet{Taniguchi23} identified hot cores in all 30 fields. 

In this paper, we present spatially resolved observations of 1.33 mm continuum toward 23 high-mass cluster-forming regions. From the DIHCA survey, for the analysis of this work, we have removed 7 fields from the original sample of 30 HMCF regions. Four of these seven targets are too close and three are heavily contaminated by free-free emission. Our observations reveal low-mass dense cores, as well as envelope and disk-like emission from young low-mass protostars, based on two ALMA datasets sensitive to different spatial scales. We present the first systematic multiplicity study of the low-mass protostellar population in high-mass cluster-forming regions to date.

\begin{figure*}
    \centering
    \includegraphics[width=15cm]{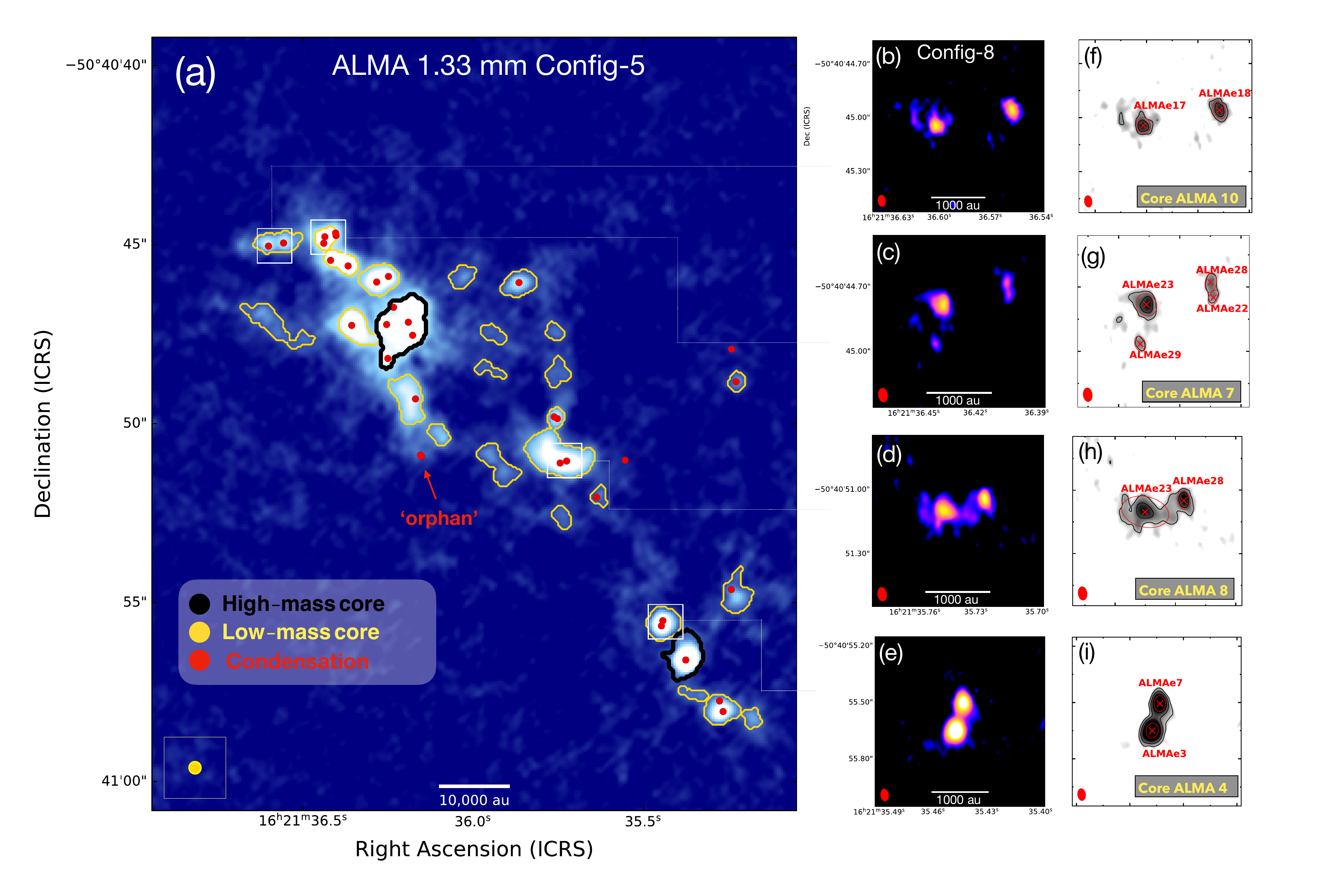}
    \caption{ The 1.33 mm continuum images of the G333.12-0.56 region. \textbf{(a)}, Background image shows the emission from 1.33\,mm dust continuum at lower resolution the synthesized beam is 0\farcs33 $\times$0\farcs28, corresponding to 1100 au at a distance of 3.3 kpc. The yellow contours outline the low-mass cores while the black contour highlights the high-mass core in this region as defined by \texttt{Astrodendro}. This image reveals 26 cores, each containing varying numbers of condensations. Condensations situated outside cores are classified as part of the orphan population. \textbf{(b)-(e)}, High-angular-resolution images of at the same wavelength showing the multiple systems embedded within the cores (beam = 0\farcs064 $\times$0\farcs041), corrsponding to 210 $\times$ 135 au. One orphan example is indicated by the red arrow. \textbf{(f)-(k)} show the same images as (b)-(e), with Gaussian fitting results along with the names of the condensations and their associated cores. Black contour shows 5$\sigma$ and 10$\sigma$ levels, where $\sigma$ = 47.2 $\mu$Jy beam$^{-1}$ is the rms noise for config-8 data. Red ellipse and cross show the condensation size and locations as defined by \texttt{PyBDSF}.}
     
    \label{1.3mm}
\end{figure*}

\section{Observations}\label{sec:obs}

We observed 23 high-mass cluster-forming regions with ALMA between 2016/11/01 and 2018/11/25 in the C43-5 configuration, and between 2016/09/11 and 2019/11/15 in the C43-8 configuration, as part of the DIHCA  project \citep{Olguin21,Olguin22,Olguin23,2025Olguin,Taniguchi23,Li24,Ishihara24,Sakai25}. The number of 12 m antennas used ranges from 41 to 46, sampling baselines up to 8500 m (Project IDs: 2016.1.01036.S, 2017.1.00237.S; PI: Patricio Sanhueza). 

We used the Common Astronomy Software Application (CASA; \citep{2007McMullin,20224CASATeam}) reduction pipeline to calibrate the data following the standard procedure. Since the data was obtained and assessed by the observatory in different cycles, each dataset was calibrated with its appropriate version of the CASA pipeline (versions 4.7.0, 4.7.2, 5.1.1-5, 5.4.0-70, and 5.6.1-8).
The 1.3\,mm continuum data is imaged using \textit{TCLEAN} task of CASA. 
The line-free channels were identified by constructing dirty data cubes for each spectral window, extracting spectra at the averaged peak position, and applying asymmetric sigma clipping to remove channels affected by line emission, as detailed in \citep{Olguin21}. 

We conducted phase and amplitude self-calibration to improve image quality. We use the Hogbom deconvolution algorithm with robust parameter Briggs weighting set to 0.5 for imaging.  
The average beam size for the 1.33\,mm continuum (at 226 GHz) is 0\farcs051 and 0\farcs32 for the C43-8 and C43-5 configurations, respectively. The average spatial resolution achieved is 960 au for C43-5, and 160 au for C43-8. 
Table \ref{tab1} summarizes the ALMA observational parameters. For the 23 high-mass cluster-forming regions, the average root mean square (RMS) noise sensitivity is 179 $\mu$Jy beam$^{-1}$ for the C43-5 data, and 42.7 $\mu$Jy beam$^{-1}$ for the C43-8 data. 

\section{Results}\label{sec:res}

The goals of this paper are (1) to determine the properties of the low-mass multiple systems embedded in high-mass cluster-forming regions to compare with those in nearby low-mass regions and (2) to assess whether turbulent core fragmentation plays a role in the formation of low-mass multiple systems embedded in high-mass cluster-forming regions. 

\subsection{ Identification of Cores and Condensations}


We have conducted high-angular resolution observations (0\farcs05 and 0\farcs3) of those cluster-forming regions in dust continuum emission at 1.33 mm using two ALMA configurations. We use mm wavelength observations because the large column of dust in high-mass cluster-forming regions precludes the direct detection of deeply embedded stellar embryos at shorter wavelengths (e.g., IR, optical, UV). At 1.33 mm and high resolution, we expect to detect the envelope and disk-like emission from young protostars. 
The compact configuration (0\farcs3) captures spatial scales that allow us to define entities typically referred to as `cores', that will form a single, a binary, or a higher-order multiple stellar system \citep{Ishihara24}.  In contrast, the extended configuration (0\farcs05) provides the angular resolution to reveal the fragmentation within the core, tracing the cocoons that will ideally form single stars \citep[referred here as `condensations',][]{Pineda15}.

Figure \ref{1.3mm} shows the G333.12-0.56 region, a typical example of the observed fields.  
This region shows a diversity of fragmentation styles that is observed across all regions: some cores condense into single or fragment into a few condensations, while others contain no embedded compact objects. In this region, certain cores host multiple condensations with projected separations ranging from 300 to 1000 au, which we classify as multiple systems. Another distinct feature is that not all condensations are associated with a dust continuum core. We refer to this last population as `orphans'.

With an average spatial resolution of $\sim$1000 au, the compact ALMA configuration reveals a total of 415 cores that are very likely formed through thermal Jeans fragmentation \citep{Ishihara24}. We therefore use the core properties obtained in Appendix~\ref{sec:CI} and \ref{sec:CM} to separate the sample into high-mass and low-mass cores. We define a high-mass core as any core with mass higher than 16 M$_{\odot}$ or cores that have characteristic `hot core' emission lines, indicative of hosting a high-mass young stellar object. 
Assuming a star formation efficiency of 50\% \citep{Matzner20}, this corresponds to a threshold core mass of 16\,M$_{\odot}$, which we adopt as the dividing line between low- and high-mass cores.
The identification of hot cores is based on \citet{Taniguchi23}, and these cores are not considered in this work. The subsequent analysis focuses on the cores that will likely form only low-mass stars ($<$8 M$_{\odot}$). With an average spatial resolution of 160 au, the newly presented high-resolution images reveal 446 condensations, either embedded within low-mass cores or located outside any identified core (See Appendix \ref{sec:CI2}).

\subsection{ Identification of Multiple Systems}

We combine two methods to define multiple systems. First, we search for multiple systems within cores, adopting the core boundary as the leaf structure decided by Astrodendro. The obtained core sizes are consistent with those from previous surveys of high-mass star-forming regions \citep{Motte22}. Using the core boundary to search for companions/multiples will increase the likelihood of identifying gravitationally bound members. Adopting the core sizes normally observed in low-mass star-forming regions (0.1 pc) is unfeasible in the clustered environments observed in high-mass star-forming regions. This is because, in our sample, the large number of cores within 0.1 pc results in an increased likelihood of chance alignments, where unrelated condensations overlap along the line of sight and are mistakenly identified as multiple systems, as we have confirmed in one of the observed regions in \citet{Li24}. Second, for orphans (condensations without a parent core), we use the largest astrodendro leaf in each region to define an area as boundary to search for companions (details are described in Appendix \ref{sec:IMS}). 
Here, a companion is defined as any other condensation located within the same core, and within a defined area in the case of orphans. For the first time, we report the discovery of 72 low-mass multiple systems in high-mass cluster-forming regions.

To ensure that our results are independent of the choice of size definition, we perform an additional analysis using a fixed size of 3000 au diameter centered on the core position. This value was chosen to avoid potential contamination from neighboring cores, as discussed in Appendix \ref{sec:CCS}. The results obtained from this approach are broadly consistent with those derived from the core boundary. 
We discuss here the statistics obtained using the core area. To address line-of-sight contamination in our sample of multiple systems, we calculate the likelihood of chance alignments for each pair and incorporate it into the analysis (See Appendix \ref{sec:MCA} and Figure \ref{fig5}). From the detected condensation population, 238 condensations (53\%) are embedded in 172 low-mass cores, while 208 condensations (47\%) are found to be unassociated with any continuum cores (orphans). Among them, 161 condensations belong to multiple systems.\\

\begin{figure*}[h]
\centering
\includegraphics[width=1\textwidth]{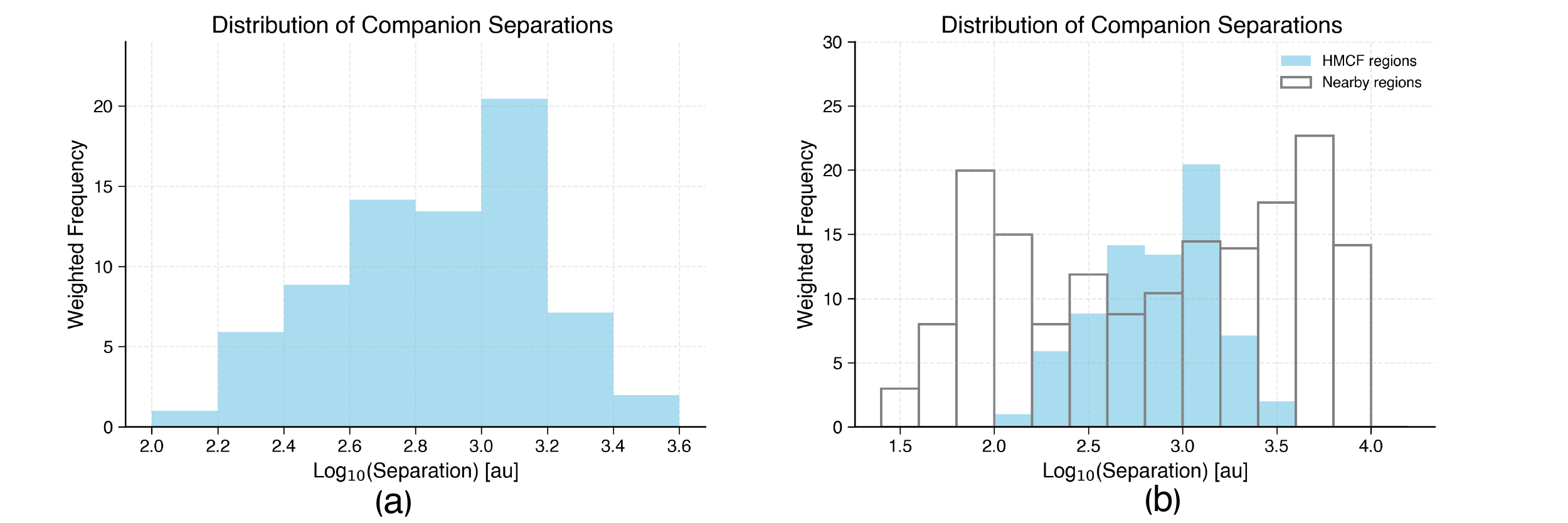}
\caption{This histogram presents companion separations for all condensations in 23 HMCF regions. Panel (a) shows results obtained using leaf-defined core.
Panel (b) presents the same results as panel (a) while also incorporating companion separations from nearby clouds, including Orion and Perseus.}
\label{fig2}
\end{figure*}

\subsection{Multiplicity}
Two informative metrics to characterize the multiplicity in star-forming regions are the multiplicity fraction (MF), which can be understood as the fraction of systems that are multiples, and the companion fraction (CF), understood as the fraction of companions per system (see details in Appendix \ref{sec:MM2}).
We identified 52 multiple systems within the cores and 20 additional multiple systems among the orphan population. Including both cores and orphans, the overall distribution of systems is S:B:T:Q = 285:59:9:4. S, B, T, and Q represent single, binary, triple, and quadruple systems, respectively. For all the low-mass condensation population in high-mass cluster-forming regions, we obtain a MF and CF of 20\% $\pm$ 2\% and 25\% $\pm$ 2\% (Calculation seen in Appendix \ref{sec:MM}).
Condensations embedded in cores have a MF and a CF of 30\% $\pm$ 3\% and 38\% $\pm$ 3\%, respectively. Orphans are primarily single-systems with a MF and a CF of 10\% $\pm$ 2\% and 12\% $\pm$ 2\%, respectively. 
In the nearby, well-studied low-mass star-forming Perseus and Orion (including the northern Integral-Shaped
Filament (ISF), L1641 and Orion B) clouds, previous surveys report a MF of about 30\% with a resolution of 30 au \citep{Tobin22}. To make a fair comparison with our survey, we recalculated their MF by treating systems with separations smaller than 160 au as a single system, corresponding to our average spatial resolution. The resulting MF in Perseus and Orion cloud is measured to be 17\% $\pm$ 2\% and 21\% $\pm$ 2\%, values that are consistent with those we obtain in high-mass cluster-forming regions.

Another useful metric is the companion separation distribution, which provides insights into the formation mechanism of multiple systems.
We adopt the method outlined by \citet{Tobin22} to calculate the companion separation for multiple systems within cores, and to identify multiple systems in the orphan population (See Appendix \ref{sec:CCS}). 
Figure~\ref{fig2}(a) shows the companion separation distribution from all multiple systems in all regions, including condensations in both within the cores and the orphan population. The distribution exhibits an asymmetric Gaussian-like shape, with a peak at $\sim$1200 au, a gradual decline as separations become smaller, and a steeper slope at larger separations. \\

\section{Discussion}\label{sec:dis}
\subsection{Physical Mechanisms of Core Fragmentation}
\begin{figure*}[h]
\centering
\includegraphics[width=0.6\textwidth]{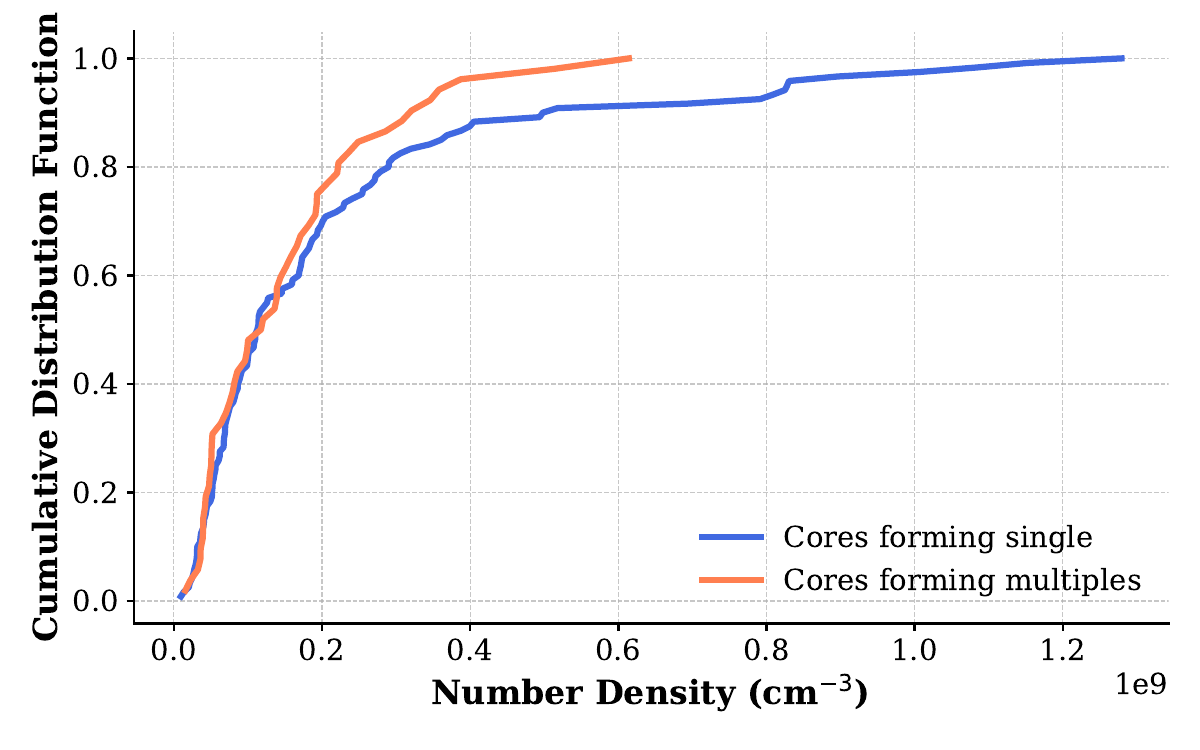}
\caption{Image shows the cumulative distribution function (CDF) of core number density, with the blue line representing cores forming single systems and the orange line representing cores forming multiple systems.}
\label{fig4}
\end{figure*}

Thermal Jeans fragmentation has been suggested to drive clump ($\sim$1 pc) to core (few 1000s au) fragmentation in the high-mass cluster-forming regions under study \citep{Ishihara24}. We examine whether it also plays a role in the fragmentation process leading to condensations (few 100s au). We assess this point following the work of \citet{Luo22}. They separate cores in Orion into two groups, those that fragment into multiples and those that do not. They show that the number density of the two groups are different, higher for cores that do fragment, in agreement with Jeans fragmentation (a higher density reduces the Jeans mass, promoting fragmentation). Following the same approach, Figure \ref{fig4} compares the number density of cores forming single and multiple systems. The Kolmogorov-Smirnov (KS) test results show the p-value is 77$\%$, which indicates no significant difference between the number density distributions of the two groups. This result also implies that factors beyond thermal Jeans fragmentation, such as turbulence and/or magnetic fields \citep[e.g.,][]{Sanhueza25}, play a more important role in the formation of low-mass multiple systems in high-mass cluster-forming regions. \\

\subsection{Fragmentation Scale in High-mass Cluster-forming Regions}
Surveys of nearby molecular clouds have revealed a bimodal companion separation distribution with one peak below 100 au and another around 4000 au \citep{Tobin22}. These two peaks are widely interpreted as the imprint of distinct formation mechanisms, although this interpretation has been challenged by simulations \citep{kh23}. The closer peak ($\textless$100 au) presumably arises from disk fragmentation caused by gravitational instabilities, while the larger peak reflects turbulent core fragmentation on scales of a few 1000s au within a collapsing core. The peak at $\sim$1200 au found in high-mass cluster-forming regions falls in between the two peaks observed in low-mass, nearby regions (Figure~\ref{fig2} (b)). Class 0 and Class I protostellar disks, as the kind of disks we would expect to find in our sample of low-mass condensations, have typical sizes smaller than 100 au \citep{Tobin20}, unresolved at the resolution of our observations \citep{Busquet19}. Furthermore, migration generally occurs inward, resulting in the contraction of member orbits \citep{Offner23,kh23}. We can therefore discard disk fragmentation as the physical process producing a peak at $\sim$1200 au. This peak is likely equivalent to the $\sim$4000 au peak in low-mass regions produced by turbulent core fragmentation, but modified by the different environmental conditions at which high-mass stars form (Appendix \ref{sec:IFS}). We propose that the turbulent and denser environment of high-mass cluster-forming regions makes the most important difference.

\begin{figure*}[h]
\centering
\includegraphics[width=0.8\textwidth]{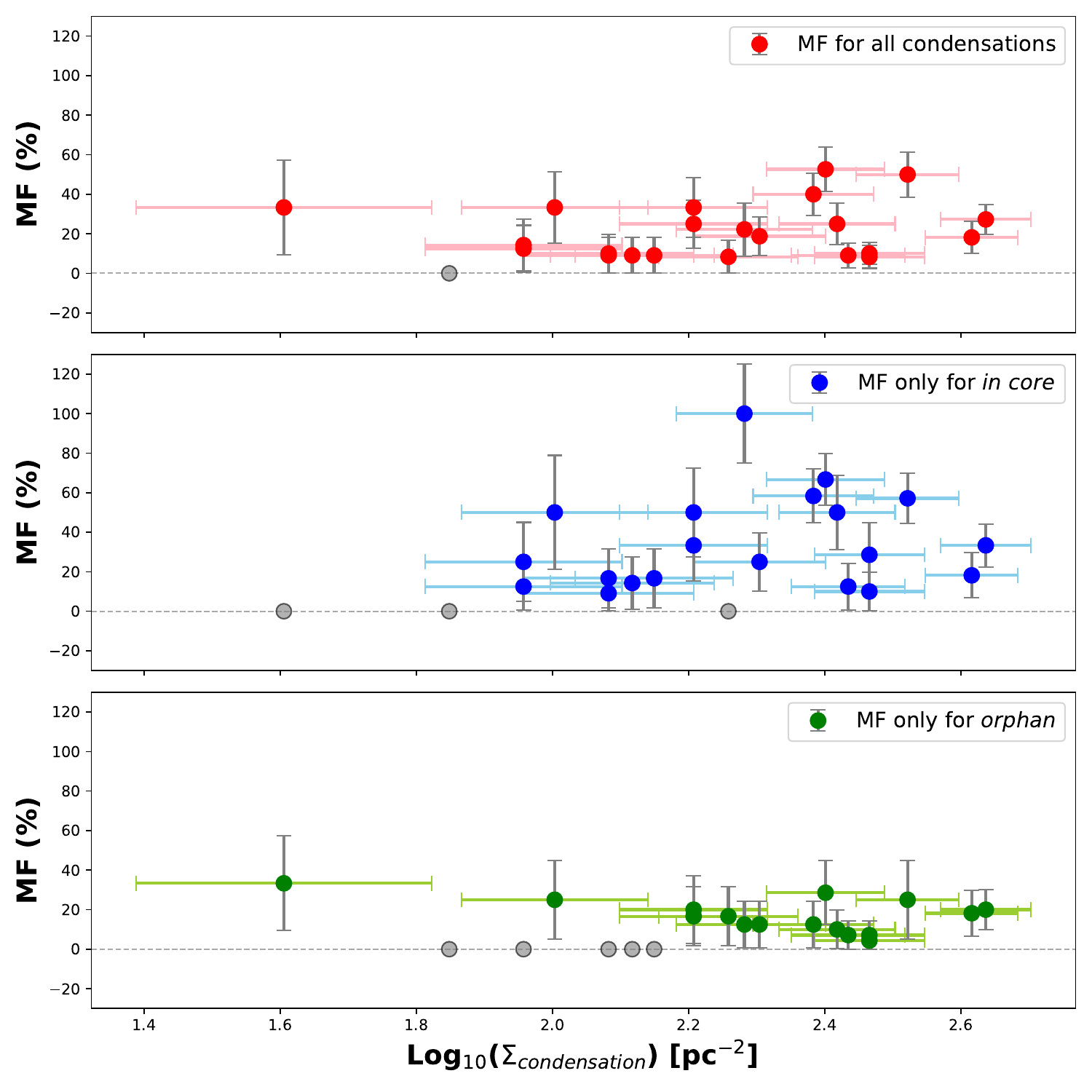}
\caption{
The figures illustrate the correlation between condensation surface density and multiplicity fraction in 23 HMCF regions. Colored dots represent the MF value points, while gray bars indicate error margins. The horizontal error bars show condensation surface density uncertainty is dominated by Poisson noise from counting objects. The first figure shows the combined multiplicity fraction results for all condensations, including both `in core' and `orphan' populations. The subsequent two figures present the same correlation separately for the in core and orphan populations. The gray filled circles indicate the region with MF = 0.
}
\label{fig3}
\end{figure*}

To investigate what defines the peak in the companion separation distribution, we first compare our observational results with numerical simulations. Self-gravitating radiation-hydrodynamics simulations  \citep{Offner10, Lee19} show that multiple systems formed via turbulent core fragmentation in clustered environments tend to have separations concentrated around 1000 au. Using magneto-hydrodynamics simulations, \citep{kh23} explore how the initial separation of systems varies with mass/density of the hosting molecular cloud. These works find that the fragmentation scale appears to be smaller in denser environments and the peak of the initial separation distribution also has a maximum near 1000 au in the simulations with the highest density.  Therefore, simulations provide further evidence that the peak of the separation distribution at $\sim$1200 au is produced by core fragmentation.\\

To examine how the star-forming environment impacts the initial fragmentation separation, we assess the contribution of turbulence by defining a fragmentation scale via a Bonnor-Ebert analysis \citep[following][]{kh23}. Rather than considering only thermal pressure and gravity (Jeans fragmentation), the Bonnor-Ebert radius of a core takes into account the pressure exerted by the turbulence and mass/density of the surrounding medium. Assuming that the typical radius at which a core fragments to produce multiples corresponding to this radius, we estimate a fragmentation scale within 600 $\pm$ 300 au and 1000 $\pm$ 500 au (See Appendix \ref{sec:IFS}), consistent with the observed peak of 1200 au from the observed companion separation distribution. We therefore propose that most of the low-mass multiple systems formed in these high-mass cluster-forming regions have been caught at an early moment in their evolution, near the moment of core fragmentation. 

Simulations suggest \citep{Offner10,Lee19,kh23} that orbital migration can quickly occur in the order of 10$^5$ yr during the embedded stage. Migration inward is predicted to reduce the separation between multiple systems due to dynamical friction that dissipates energy and angular momentum \citep{kh23}. This process is particularly effective when the forming stars are not yet sufficiently massive with respect to the surrounding medium, and gas dynamics dominate. We interpret the asymmetry in the companion separation distribution, specifically the shallow slope toward smaller separations, as evidence for orbital migration \citep{kh23}. On the other hand, the steeper slope toward larger separations reflects the compactness of high-mass cluster formation \citep{Guszejnov23}. The sharp drop in the distribution naturally arises from restricting the search of multiples to within the boundary of individual cores (or within 3000 au). If this limit is relaxed, in addition to increasing the likelihood of chance alignments, the association of multiples becomes contaminated by objects belonging to neighboring cores (See Appendix Figure \ref{fig5}(b)).

\subsection{Condensation Surface Density Effects}

Multiplicity variations in nearby low-mass star-forming regions may largely reflect differences in stellar surface densities, which in turn could promote multiple system formation and lead to a higher MF \citep{Tobin22,Offner23}. We investigate this effect in our sample of condensations embedded in high-mass cluster-forming regions and find no correlation. Figure~\ref{fig3} (also Table~\ref{tab2}) presents the condensation MF as a function of surface density, ranging from 40 to 433 pc$^{-2}$. For comparison, the average surface density or condensations, as measured in Perseus and Orion, is 30 pc$^{-2}$ \citep{Tobin22}, whereas the average surface density for condensations in this study of high-mass regions is 210 pc$^{-2}$. Figure~\ref{fig3} illustrates how the MF remains relatively constant across different condensation surface densities, regardless of whether all condensations are included or the sample is divided into those embedded in cores and orphans. The discrepancy between low-mass and high-mass regions may suggest that the condensation surface density plays a different role in shaping multiplicity. We propose that the more chaotic and dynamical conditions prevalent in high-mass environments disrupt the tendency observed in low-mass regions. At sufficiently high surface densities, these conditions may hinder the ability to retain binary systems, particularly those that are less gravitationally bound. This interpretation aligns with the cluster-forming simulations conducted by \citep{Guszejnov23}, which show that multiplicity decreases with increasing local stellar density. Indeed, the discovery of the orphan population provides further evidence for the highly dynamical conditions in high-mass cluster-forming regions. Approximately 47\% of the orphans are co-located with diffuse continuum emission, but are not within a core. This subpopulation may have been ejected from cores due to dynamical interactions. On the other hand, the remaining 53\% of the orphan population is unassociated with diffuse dust continuum and appears to be more isolated. This subpopulation of orphans may be an older generation of protostars formed in low-mass cores that have already dispersed most of their envelope. As shown in Figure~\ref{fig6} and detailed in the Appendix \ref{sec:OP}, as a proxy for the condensation mass in the optically thin regime, the millimeter luminosity of the orphan population is generally lower than that of condensations embedded in cores. With both, a lower mass distribution and a lower MF of 8\%, the orphan population has properties consistent with the lowest-mass stars ($\sim$0.1 M$_{\odot}$) in the Main Sequence  phase \citep{Offner23}. 
\\

\section{Conclusions}\label{sec:sum}
We provide, for the first time, direct observational evidence from a statistically significant sample that turbulent core fragmentation plays a role in determining multiplicity in the high-density environments where high-mass stars form. The larger turbulence and high densities observed in high-mass cluster-forming regions alter the characteristic fragmentation scale compared to that produced by core fragmentation in low-mass regions. Despite having larger condensation surface densities, the MF in high-mass cluster-forming regions does not increase. We suggest that the shallow slope into companion separation distribution at small separations is produced by inward migration, as suggested by simulations, while the steep slope at larger separations results from high clustering. The newly identified orphan population is notable for its significant numbers but predominantly forms single stars, a feature that may be linked to the cluster formation process. The strong dynamical interactions in these highly dense clusters likely lead to high-order gravitational interactions, hindering the formation of weakly bound binaries and ejecting a large population of objects, which likely contributes to a large portion of the orphan population.
\\

\begin{acknowledgments}
QY-L acknowledges the support by JSPS KAKENHI Grant Number JP23K20035.
PS was partially supported by a Grant-in-Aid for Scientific Research (KAKENHI Number JP22H01271 and JP23H01221) of JSPS. 
This paper makes use of the following ALMA data: ADS/JAO.ALMA\#2016.1.01036.S, ADS/JAO.ALMA\#2017.1.00237.S, 
ADS/JAO.ALMA \#2017.1.00101.S, and ADS/JAO.ALMA\#2018.1.00105.S. ALMA is a partnership of ESO (representing its member states), NSF (USA) and NINS (Japan), together with NRC (Canada), MOST and ASIAA (Taiwan), and KASI (Republic of Korea), in cooperation with the Republic of Chile. The Joint ALMA Observatory is operated by ESO, AUI/NRAO and NAOJ.
Data analysis was in part carried out on the Multi-wavelength Data Analysis System operated by the Astronomy Data Center (ADC), National Astronomical Observatory of Japan.
SSR Offner acknowledges support from NSF AAG 2407522.
F.O. acknowledges the support of the NAOJ ALMA Joint Scientific Research Program grant No. 2024-27B.
S.L. acknowledges support from the National SKA Program of China with No. 2025SKA0140100, “Double First-Class” Funding with No. 14912217, and National Natural Science Foundation of China (NSFC) grant with No. 13004007. 
K.E.I.T.  acknowledges the support by JSPS KAKENHI grant No. 25K07365 and NAOJ ALMA Scientific Research Grant Code 2026-31B.
Y.C. was partially supported by a Grant-in-Aid for Scientific Research (KAKENHI  number JP24K17103) of the JSPS

\end{acknowledgments}




%
\facilities{ALMA}




\appendix

\section{Source Extraction}\label{sec:SE}
\subsection{Core Identification}\label{sec:CI}

Using \texttt{Astrodendro} \footnote{https://dendrograms.readthedocs.io/en/stable/} Package, \citet{Ishihara24} identify compact structures in the 1.33\,mm dust continuum emission using the C43-5 configuration data (0\farcs3). In their work, the `leaf', the highest structure in the hierarchy tree, is defined as the `core'. 
The parameters obtained from \citet{Ishihara24} include coordinates, flux density, core radius, and core area for all identified structures, as provided by \texttt{Astrodendro}. These results are summarized in the Table \ref{tab3}.
A total of 415 cores are identified among the 23 regions studied, containing a mixed population that includes high-mass cores (including hot cores), and low-mass cores. 

Fixed-size `cores': 
To facilitate a comparison with nearby cloud statistics, where the core size remains relatively constant, we also adopt a uniform core size.
A constant core size also allows us to assess the effect in the analysis of having cores with different sizes. However, the maximum size that could be adopted is restricted by the proximity of cores in the clustered environments under study. If the adopted size is excessively large, we will combine independent cores with their respective multiple systems. To explore what fixed core size could be used, we run the algorithm defining multiples up to 6000 au, ignoring core boundaries. 
Figure \ref{fig5} presents a comparison of the condensation companion separation distribution (obtained from Appendix \ref{sec:CCS}) and the core separation distribution (obtained from \citet{Ishihara24}), revealing an overlap between the two distributions. The results are only used to test whether the multiplicity statistics are sensitive to the choice of core size, and they are shown only in Figure \ref{fig5}.
To mitigate contamination in the definition of multiple systems caused by condensations from nearby cores, we apply a fixed core size of 3000 au to all cores in the subsequent multiplicity analysis. The fixed size of 3000 au reduces the contamination from neighboring cores to 2$\%$.

\subsection{Condensation Identification}\label{sec:CI2}
Using the \texttt{PyBDSF}\footnote{https://pybdsf.readthedocs.io/en/latest/} package, we identify `condensations' and their physical properties using the C43-8 configuration data (0\farcs05). 
\texttt{Astrodendro} is well-suited for extracting cores with extended structures in lower-resolution data. 
As the condensations we aim to extract are predominantly point-like sources in the extended configuration, achieving higher precision is essential. \texttt{PYBDSF} employs an adaptive scaling approach to identify condensations across a wider range of emission scales, enabling accurate capture of point-like sources. 
This method is well suited for our long-baseline data. The detection parameters are listed in Table \ref{tabp}.

We then performed a detailed visual inspection of each cataloged condensation to ensure its reliability. By analyzing their dust emission morphology and assessing the Gaussian fitting results, we exclude condensations that are likely interferometric artifacts or over-fitting results. 
After filtering, our results reveal a total of 536  condensations , including all those embedded in low- and high-mass cores in the 23 regions. The condensation catalog has been filtered to include only those that either formed within low-mass cores or are not associated with any core, result in 446 are condensations. Table \ref{tab4} lists their coordinates, deconvolved radii (defined as half of the FWHM), and total fluxes.

\section{Core Property Measurements}\label{sec:MM}

\subsection{Core Mass}\label{sec:CM}
We use the 1.33\,mm dust emission to estimate the core mass. Assuming that the dust emission from cores is optically thin and isothermal, and that the gas and dust are well mixed, we can calculate the mass of a core from the expression:

\begin{equation}
    M_{core}=\mathbb{R}~\frac{S_{\nu}D^2}{\kappa_{\nu}B_{\nu}(T_{\rm dust})}~.
\end{equation}

Here, S$_{\nu}$ represents the flux density of the cores, $D$ is the distance to each region, $\mathbb{R}$ is the gas-to-dust mass ratio of 100, and the dust opacity $\kappa_{\nu}$=0.899 cm$^2$ g$^{-1}$ corresponds to the opacity of dust grains with thin ice mantles at a gas density of $10^6$ cm$^{-3}$ \citep{Ossenkopf94}. $B_{\nu}$ is is the Planck function for a blackbody at the dust temperature $T_{\rm dust}$.
We adopt the hot core temperature determined by \citep{Taniguchi23} as the dust temperatures when available. For the remaining cores within a given region, we adopt the single-dish dust temperature reported by \citet{Ishihara24}. 
The distance and dust temperature information in Table \ref{tab2}. 

The masses of the identified cores range from 0.071 to 33.5 M$_{\odot}$. To separate the core sample between high-mass and low-mass, we first assumed that all hot cores identified by \citet{Taniguchi23} will likely form a high-mass star, independent of their masses. Cores with masses exceeding 16 M$_{\odot}$ are classified as high-mass cores. The remaining ones, with masses below 16 M$_{\odot}$ (excluding hot cores), are classified as low-mass cores. These cores have a mass ranging from 0.071 to 13.97 M$_{\odot}$. As a result, we have 378 low-mass cores and  37 high-mass cores in the 23 regions.

\subsection{H$_2$ Number Density of Cores}\label{sec:PPC}

We calculate the H$_2$ number density assuming a spherical core as follows,
\begin{equation}
    n_{\rm core} = \frac{M_{\rm core}}{(4/3)\pi R_{\rm core}^3 \mu m_H}~.
\end{equation}

Here, $M_{\text{core}}$ represents the core mass, and $R_{\rm core}$ is the core radius. $\mu$ = 2.8 \citep{2008Kauffmann} is the  mean molecular weight, and $m_H$ is the mass of a hydrogen atom. The number density of cores is calculated and grouped into two categories: one containing cores that formed single systems, and another one containing cores that formed multiple systems. 
The cores forming single systems have number densities ranging from $9.6 \times 10^{6} \,\mathrm{cm^{-3}}$ to $1.3 \times 10^{9} \,\mathrm{cm^{-3}}$, with a median value of $1.1 \times 10^{8} \,\mathrm{cm^{-3}}$. The cores forming multiple systems have number densities ranging from $1.6 \times 10^{7} \,\mathrm{cm^{-3}}$ to $6.4 \times 10^{8} \,\mathrm{cm^{-3}}$, with a median value of $1.2 \times 10^{8} \,\mathrm{cm^{-3}}$.
We perform a Kolmogorov-Smirnov (KS) test on the cumulative distribution function of the number density for both core categories. A p-value smaller than 0.05 would indicate that the two distributions differ significantly. The KS test yields  a p-value of 0.77 and a statistic of 0.11, indicating no statistically significant difference between the two distributions (See Figure \ref{fig4}). 
Detailed information, including core ID, coordinates, radius, total flux and number density for all low-mass cores, is presented in the Table \ref{tab3}.\\
We note that the uncertainty in the mass and the density is dominated by the uncertainty of the gas-to-dust mass ratio and the dust opacity, with 23\% and 28\% being the respective values \citep{Sanhueza17,Sanhueza19}, resulting in a mass and number density uncertainty of about 50\% their value.

\section{Orphan population}\label{sec:OP}
Region-based matching was carried out using the spatial extent of each core to identify condensations located within its boundary (Appendix \ref{sec:IMS}). Because in this work we focus on the multiplicity of low-mass, newly forming stars in high-mass cluster-forming regions, we exclude high-mass and hot cores together with their corresponding condensations. The remaining condensations (embedded in low-mass cores or without a host core) are considered more likely to evolve into low-mass stars.
 Our analysis reveals that 237 (53$\%$) from the whole population of condensations are associated with low-mass cores, whereas 208 (47$\%$) are located outside core regions. For this 47$\%$, referred to as the `orphan' population, we tentatively propose two possible scenarios to explain their origin based on where the condensations are located. 
In the first scenario, around 94 (45$\%$) of orphans are found to be associated with diffuse emission in the lower-resolution data, but are not associated with any core. These orphans are, however, located in close proximity to cores. We suggest that they probably form within a core, but N-body interactions resulting from multiple system formation led to the ejection of the most unbound member of a system. While a fraction of the orphan population may originate from dynamical ejection, such sources are not necessarily isolated permanently, as they may later be bound in new multiple systems through dynamical capture.
In the second scenario, the remaining 114 (55$\%$) of orphans are clearly isolated, showing no detectable association with surrounding dense structures. These orphans tend to be away from dust continuum cores. They probably originated in an older generation of low-mass cores that became undetectable over time as most of the surrounding gas dissipated during their evolution.

Figure~\ref{fig6} compares the distribution of mm luminosity (flux $\times$ distance square)  of `in core' and `orphan' populations. 
The distribution of mm luminosity of the orphan population shows a median value of 4 mJy kpc$^2$, which is significantly lower than the median of 19 mJy kpc$^2$ for condensations in low-mass cores. 
Although the two distributions partially overlap, the orphan population shows a significantly larger proportion at lower mm luminosity, suggesting a clear distinction in their overall mm luminosity ranges. 
Assuming that the mm luminosity is proportional to the condensation mass, under optically thin conditions, this difference may suggest the discontinuation of accretion due to ejection and/or by dissipation of the parent core. These findings imply that both origin scenarios for the orphan population coexist, contributing jointly to the observed characteristics of this population. 

\section{ Identifying Multiple System}\label{sec:IMS}

We identify multiple systems separately in the cores and orphan population. 
First, multiple systems in cores are identified by cross-matching the regions of core and condensations. Cores containing two or more condensations within their boundaries are classified as multiple systems, while those with only one condensation are classified as single systems.
Secondly, for the orphan population, we define the effective radius as $Area = \pi R_{\rm eff}^2$ ($Area$ defined by the Astrodendro leaf) and adopt 2 times the largest $R_{\rm eff}$ in the field for searching for multiples. We use 2 times because it allows the multiples to be at the edge of a circle of a diameter equal to 2$R_{\rm eff}$. The largest $R_{\rm eff}$ values are listed in Table  \ref{tab2}.
In total, we identify 72 low-mass multiple protostellar systems in the 23 observed regions. Of these systems, fifteen multiple systems come from the orphan population. 
Table \ref{tabpp} includes all pairings of the 72 systems and shows each is located within a core or in an orphan group.

\section{ Measuring Multiplicity}\label{sec:MM2}

To gather the multiplicity statistics of low-mass condensations in high-mass cluster-forming regions, we adopt the formulation from \citep{Tobin22} to calculate two key parameters, the multiplicity fraction (MF) and the companion fraction (CF), along with their uncertainties, as shown below: 
\begin{equation}
    MF=\frac{B + T + Q+... }{S + B + T + Q+...}
\end{equation}

\begin{equation}
    CF = \frac{B + 2T + 3Q+... }{S + B + T + Q+...}
\end{equation}

\begin{equation}
\sigma_{\text{MF}} = 
\frac{1}{1 + \frac{z^2}{N_{\text{sys}}}}
\left(
\text{MF} + \frac{z^2}{2N_{\text{sys}}} \pm z \sqrt{
\frac{\text{MF}(1-\text{MF})}{N_{\text{sys}}} + \frac{z^2}{4N_{\text{sys}}^2}
}
\right),
\end{equation}

\begin{equation}
\sigma_{\text{CF}} = 
\frac{1}{1 + \frac{z^2}{N_{\text{sys}}}}
\left(
\text{CF} + \frac{z^2}{2N_{\text{sys}}} \pm z \sqrt{
\frac{\text{CF}(1-\text{CF})}{N_{\text{sys}}} + \frac{z^2}{4N_{\text{sys}}^2}
}
\right),
\end{equation}

Here, \textit{S}, \textit{B}, \textit{T}, and \textit{Q} are the numbers of single, binary, triple, and quadruple systems in the entire sample, respectively. $N_{\text{sys}}$ is the total number of systems (e.g., $S + B + T + Q + \ldots$). The parameter $z$ corresponds to the standard normal quantile, which is set to $z=1$ to represent the $\sigma_{MF}$ and $\sigma_{CF}$ uncertainties. 

Within all low-mass cores that contain condensations (52), the distribution of protostellar systems is S:B:T:Q = 120:42:6:4. In the orphan population, the distribution is S:B:T:Q = 165:17:3:0.
Furthermore, we calculate the MF for each region under three scenarios: considering all condensations, only those within cores, and those in the orphan population. The detailed results are provided in Table \ref{tab2}.

\section{ Calculating Companion Separation }\label{sec:CCS}

To analyze the separation between member condensations in a multiple system, we adopt the inside-out search method from \citet{Tobin22} for a consistent comparison with previous observations and simulations. The procedure for calculating the companion separations in each cluster-forming region is as follows:
(1) The search begins from any condensation within a core, with a minimum search radius of 50 au.
(2) We search for companions for each condensation by incrementally increasing the radius in 10 au steps, extending the search up to a defined boundary.

We first calculate the companion separation for multiple systems by applying the procedure to each core, based on the condensations extracted from their respective core regions.
Subsequently, the same procedure is applied to the orphan group in each region, with the search limit set to the 2$R_{\rm eff}$ in the region and excluding condensations that are within cores.

\begin{figure*}
\centering
\includegraphics[width=1.1\textwidth]{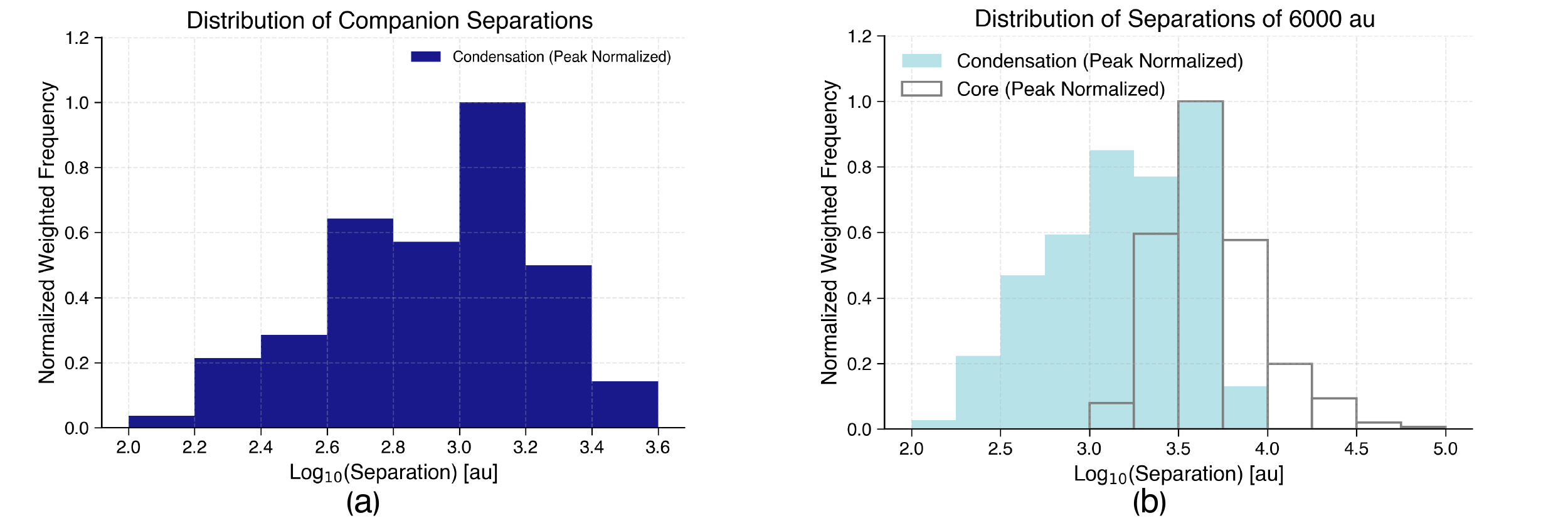}
\caption{ Panel (a) shows the companion separation distribution for a fixed core size of 3000 au in 23 HMCF regions. Panel (b) displays the companion separation distribution without any core size constraints in 23 HMCF regions, with separations limited to a maximum of 6000 au. The gray histogram represents the core separation distribution in 23 regions from \citet{Ishihara24}. To facilitate comparison, both distributions have been peak-normalized.}
\label{fig5}
\end{figure*}

\begin{figure*}
\centering
\includegraphics[width=0.6\textwidth]{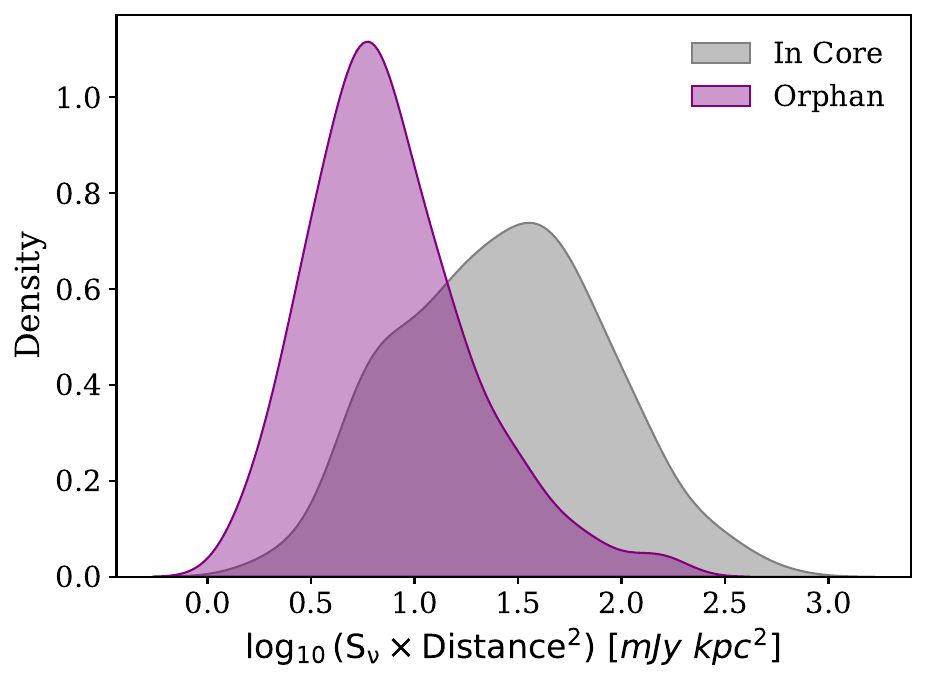}
\caption{
The figure shows the distribution of mm luminosity (flux $\times$ distance$^2$) for two groups: gray reflects condensations located in low-mass cores, while purple illustrates condensations labeled as `orphan'.}
\label{fig6}
\end{figure*}

The companion separation distributions shown in Figures \ref{fig2} and \ref{fig5} include contributions from condensations within cores (under two conditions: identified `leaf' cores and a fixed core size of 3000 au) and from the orphan population (under two conditions: 2$R_{\rm eff}$ and a fixed core size of 3000 au). 
For these histograms, each separation is weighted according to the probability of being a multiple system, 
as detailed in Appendix~\ref{sec:MCA}. 
This weighting accounts for the uncertainty in the identification of overlapping systems along the line of sight, ensuring that the distribution reflects the likelihood of multiplicity rather than treating all systems as equally probable.

To examine the effects of the core definition, we search for multiples ignoring core's boundaries. We calculate the companion separation blindly to all condensations up to 6000 au.   
Figure \ref{fig5} compares the resulting companion separation distribution with the core separation distribution derived by \citet{Ishihara24} using the minimum spanning tree (MST) algorithm. The MST algorithm connects a set of core centers using the shortest possible network of branches without forming loops. 
The companion separation distribution reveals two peaks: one at 10$^{3.1}$ au ($\sim$1200 au) and another around 10$^{3.6}$ au($\sim$4000 au). The second peak coincides with the peak of the core separation distribution from \citep{Ishihara24}, indicating that if multiples are defined blindly, without a boundary, condensations embedded in different cores will be considered as multiples. Therefore, we interpret the $\sim$1200 au peak as a robust characteristic fragmentation scale and the true companion separation.


\section{ Measuring Condensation Surface Density}\label{sec:MCSD}
To investigate how condensation surface density (which can be considered analogous to YSO surface density) affects multiplicity, we derive the surface density from the 23 observed regions.
To enable a consistent comparison across these regions, for all of them, we assumed an area equal to that of G14.22-0.50 S, the closest source, with a defined Field of View (FOV) of 38$^{\prime\prime}$ and a physical area of 0.098 pc$^2$. The angular size for other regions were scaled according to their distances. 
This FOV and area standardization approach was then consistently applied across all regions to ensure uniformity in our calculations:
\begin{equation}
    \Sigma_{condensation} = \frac{N_{condensation}}{A}~.
\end{equation}

Here, $N_{\text{condensation}}$ represents the number of condensations within a fixed area, A = $0.098\,\text{pc}^2$. The calculated condensation surface density ranges from 40 condensations per pc$^{2}$ to $433$ condensations per pc$^{2}$. Poisson uncertainties were estimated from the condensation counts and propagated accordingly. A complete list of values is presented in Table \ref{tab2}.

\section{ Measuring Chance Alignment}\label{sec:MCA}

To assess the reliability of companion separations and exclude the impact of random alignment along the line of sight, we employ the method from \citet{Tobin22} to calculate the probability of being a real companion (the complement of the probability of chance alignment) for each pair of condensation companions. 

We assume that the number of condensation chance alignment follows a Poisson distribution, given by: 
\begin{equation}
P(k) \;=\; \frac{\lambda^k e^{-\lambda}}{k!},
\end{equation}
where $\lambda$ = $\tau \Sigma\pi d^2$, and $k$ is the number of condensations expected.

\begin{figure*}
\centering
\includegraphics[width=0.6\textwidth]{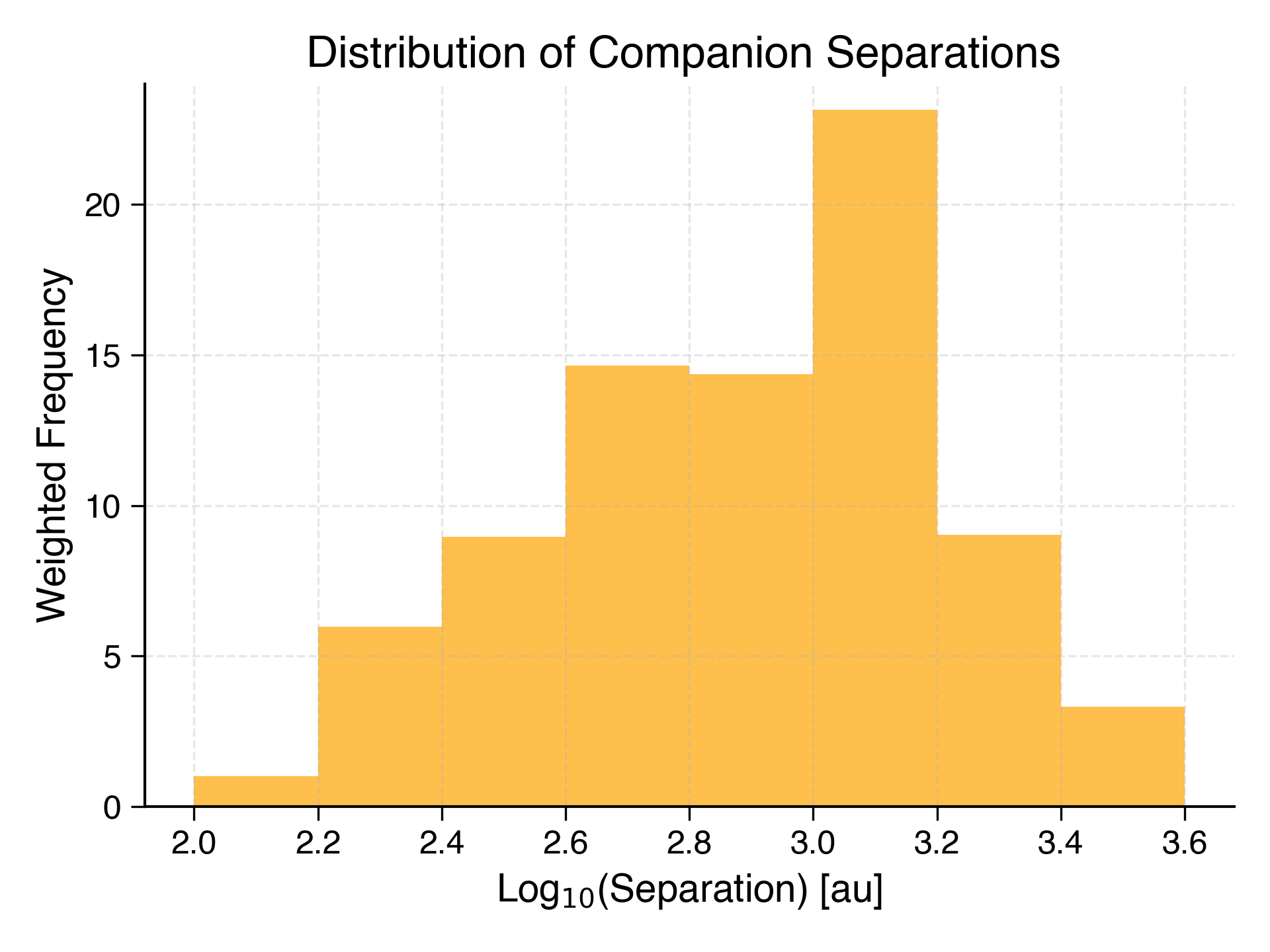}
\caption{The companion separation distributions as a function of condensation surface density $\Sigma_{cond}$ are summarized in Table \ref{tab2}.
}
\label{fig7}
\end{figure*}

Thus, the probability of detecting more than one condensation is given by 1~-~P(0), where P(0) is the Poisson distribution of non-detection of a condensation.
\\
The probability of detecting more than one condensation is given by: \\

\begin{equation}
        P(\geq 1 \,\,\text{condensation}) = 1-e^{\lambda}=1 - e^{-\tau \Sigma \pi d^2}.
\end{equation}
Here, $\tau$ = 0.4 accounts for the observational completeness, which we obtained by comparing the condensation flux density in our sample with that from \citet{Tobin22}. In their work, the completeness was estimated as 75\% based on Lupus surveys \citep{Ansdell2016}, but applying the same approach to our sample yields a lower value of 40\%.
The variable $d$ represents the separation between two condensation companions, measured in $pc$. 
For the surface density, we assume as a representative field stellar surface density the one from the ONC, $\Sigma$ = 770 pc$^{-2}$ \citep{Reipurth2007}. Additionally, in Figure~\ref{fig7}, we adopt the condensation surface density derived for each specific field (Appendix~\ref{sec:MCSD}). This companion separation distribution exhibits a similar shape and also peaks around 1000~au. 

We then use the Bayes' theorem to determine the probability that detected a source is a true companion:\\
\begin{equation}
        P(\text{companion} \mid \text{detection}) = \frac{P(\text{detection} \mid \text{companion}) \cdot P(\text{companion})}{P(\text{detection})}.
\end{equation}
Here, $P(\text{detection} \mid \text{companion})$ is the probability of detecting a companion, which we set equal to the completeness $\tau$=0.4. This assumption has a negligible impact on the result. The companion probability, $P(\text{companion})$, is simply treated as the companion fraction (CF), which is iteratively updated in our analysis.
The denominator, $P(\text{detection})$ is the total probability of detecting any condensation, which can be calculated using the following equation:\\
\begin{equation}
        P(\text{detection}) = \tau \, P(\text{companion}) + \left(1 - e^{-\tau \Sigma \pi d^2}\right) \left(1 - \tau \, P(\text{companion})\right).
\end{equation}

For each candidate pair, we define the number of expected companion in each system as follow:
\begin{equation}
E_i(CF) = \sum_{k=1}^{\min(m_i, r_i)} \Bigl(r_i - (k-1)\Bigr)
\Biggl[\prod_{j=1}^{k-1}\bigl(1-p_{i,j}(CF)\bigr)\Biggr] p_{i,k}(CF),
\end{equation}
where $p_{i,k}$(CF) is the conditional probability of the $k$-th pair being a true companion $P(\text{detection} \mid \text{companion})$. $m_i$ is the number of candidate pairs in system $i$ and $r_i$ = $n_i$ - 1 is the maximum possible number of companions for that system (with $n_i$ being the number of detected members). The index $k$ orders the candidate pairs by increasing companion separation, so that the closest pair is evaluated first. The global CF is then obtained by solving the above equation iteratively until convergence.
\begin{equation}
CF = \frac{1}{N_{\mathrm{sys}}} \sum_i E_i(CF).
\tag{I13}
\end{equation}
We calculate the probability of being a real companion  $P(\text{companion} \mid \text{detection})$ for each candidate pair and list them in Table \ref{tabpp}. These probabilities, denoted as $p_{i,k}$(CF), are used both as statistical weights in the analysis of companion separation distributions and as inputs to the chain–expectation calculation of the global CF in each region, ensuring that every pair contributes in proportion to its statistical reliability.

\section{ Initial Fragmentation Scale}\label{sec:IFS}
We apply the approach described in \citet{kh23} to estimate the initial fragmentation scale as a proxy for turbulent core fragmentation, which approximates how the star-forming environment influences the separation of multiple systems within cores considering the surrounding density and turbulence. This approach assumes that cores, on average, can be modeled as critical-mass Bonnor-Ebert spheres. The fragmentation scale is then approximated to the radius of a critical Bonnor-Ebert core modified by an efficiency factor, $\epsilon_{core}$, as

\begin{equation}
    d_{\rm core}=\epsilon_{\rm core}R_{\rm BE} = 0.485\epsilon_{\rm core}\frac{c_s^2}{G^{1/2}P^{1/2}_{\rm ext}}~,
    \label{frag_scale}
\end{equation}
in which $\epsilon$ is an efficiency factor that we consider to be between 0.55  \citep[from simulations in][]{kh23} and 1.0 (fragmentation of the core occurring at the edge of the core). 
The sound speed, $c_s$, is given by

 \begin{equation}
    c_s = \sqrt{\frac{k_{\rm B} T_{\rm gas}}{\mu\, m_{\rm H}}}~,
    \label{eq_th}
\end{equation}
in which $\mu=2.37$ is the mean molecular weight per free particle used for the sound-speed calculation, and $T_{\rm gas}$ is the clump temperature. P$_{ext}$ denotes the external pressure confining the Bonnor-Ebert sphere,

\begin{equation}
    P_{\rm ext} = (1+\mathcal{M}_s^2)P_{\rm th}=(1+\mathcal{M}_s^2)\,c_s^2 \, \rho_{\rm gas}~,
\end{equation}
in which $\rho_{\rm gas}$ is the clump density, and $\mathcal{M}_s$ is the Mach number given by $\mathcal{M}_s = \sigma_{\rm nt}/c_s$. $\sigma_{nt}$ is the non-thermal velocity dispersion derived from $\sigma_{\rm nt}$ = $\sqrt{\sigma_v^2-\sigma_{\rm th}^2}$ and $\sigma_v$ = $\frac{FWHM}{2\sqrt{2\ln2}}$ is the observed velocity dispersion, where FWHM is the full width at half maximum of HN$^{13}$C $J=3-2$ lines. The thermal velocity dispersion, $\sigma_{\rm th}$, of HN$^{13}$C molecules is calculated using Equation~\ref{eq_th} by replacing $\mu$ for the molecular weight of HN$^{13}$C, 28. We extracted the average line FWHM of the envelope surrounding each core using the optically thin HN$^{13}$C $J=3-2$ transition line from the MagMaR (Magnetic Fields in Massive Star-forming Regions) project \citep{Fernandez21,Sanhueza21,Sanhueza25,Cortes21,Cortes24,Saha24,Zapata24}, which observed the same sample of this study at the same angular resolution than the C-5 configuration ($\sim$0\farcs3). 

To evaluate the initial fragmentation scale using Equation~\ref{frag_scale} and compare it with the observed characteristic separation of the companion separation distribution, we calculate the average values of the clump density ($\rho_{\rm gas}=7.2\times10^5$ cm$^{-3}$), sound speed ($c_s=0.28$ km s$^{-1}$), and non-thermal speed ($\sigma_{\rm nt}=0.58$ km s$^{-1}$). This yields a characteristic fragmentation scale of 600-1000 au. We assess the accuracy of the calculation by accounting for the spread in the individual distributions of $\rho_{\rm gas}$, $c_s$, and $\sigma_{\rm nt}$, using their standard deviation as uncertainties. By performing a Monte Carlo simulation, we finally find that the initial fragmentation scale derived from the Bonnor-Ebert analysis, which includes the density and turbulence of the surrounding medium, ranges from 600 $\pm$ 300 to 1000 $\pm$ 500 au. This result is consistent with the observed peak of $\sim$1200 au in the companion separation distribution.

\clearpage
\input{combined_obs}

\clearpage
\input{clump_information}
\clearpage
\input{low_mass_cores_table}
\input{pybdsf-parameter}
\input{low_mass_ysos}

\input{low_mass_condensation_pairs}
\clearpage
\bibliography{main}{}
\bibliographystyle{aasjournalv7}



\end{document}

%% file: combined_obs.tex
\clearpage
\begin{sidewaystable}

\begin{longtable}{lcccccccccccc}

\caption{Summary of ALMA Observations} \label{tab1} \\
\toprule
Clump name & Dist & RMS$_\mathrm{C5}$ & RMS$_\mathrm{C8}$ & Beam$_\mathrm{C5}$ & Beam$_\mathrm{C8}$ & Res$_\mathrm{C5}$ & Res$_\mathrm{C8}$ & Baseline$_\mathrm{C5}$ & Baseline$_\mathrm{C8}$ & $N_\mathrm{ant}$,$_\mathrm{C5}$ & $N_\mathrm{ant}$,$_\mathrm{C8}$ \\
& (kpc) & ($\mu$Jy beam$^{-1}$) & ($\mu$Jy beam$^{-1}$) & ($^{\prime\prime} \times ^{\prime\prime}$) & ($^{\prime\prime} \times ^{\prime\prime}$) & (au) & (au) & (m) & (m) & & \\
(1) & (2)& (3) & (4) & (5) & (6) & (7) & (8) & (9) & (10) & (11) & (12) \\
\midrule
\endfirsthead

\multicolumn{13}{l}{\textbf{Table \thetable{} -- Continued from previous page}} \\
\toprule
Clump name & Dist & RMS$_\mathrm{C5}$ & RMS$_\mathrm{C8}$ & Beam$_\mathrm{C5}$ & Beam$_\mathrm{C8}$ & Res$_\mathrm{C5}$ & Res$_\mathrm{C8}$ & Baseline$_\mathrm{C5}$ & Baseline$_\mathrm{C8}$ & $N_\mathrm{ant}$$_\mathrm{C5}$ & $N_\mathrm{ant}$$_\mathrm{C8}$ \\
& (kpc) & ($\mu$Jy beam$^{-1}$) & ($\mu$Jy beam$^{-1}$) & ($^{\prime\prime} \times ^{\prime\prime}$) & ($^{\prime\prime} \times ^{\prime\prime}$) & (au) & (au) & (m) & (m) & & \\
(1) & (2)& (3) & (4) & (5) & (6) & (7) & (8) & (9) & (10) & (11) & (12) \\
\midrule
\endhead

\midrule
\multicolumn{13}{r}{\textit{Continued on next page}} \\
\endfoot

\bottomrule
\endlastfoot
G11.1-0.12 & 3.0 & 75 & 27.8 & 0.41$\times$0.25 ($-74.3^\circ$) & 0.057$\times$0.044 ($77.8^\circ$) & 1100 & 150 & 15.1--1100 & 92--8500 & 45 & 44 \\
G11.92-0.61 & 3.37 & 89 & 36.5 & 0.41$\times$0.25 ($-75.7^\circ$) & 0.055$\times$0.041 ($69.1^\circ$) & 970 & 160 & 15.1--1100 & 92--8500 & 50 & 44 \\
G14.22-0.50S & 1.9 & 139 & 34.5 & 0.42$\times$0.26 ($-73.6^\circ$) & 0.051$\times$0.044 ($68.1^\circ$) & 830 & 90 & 15.1--1100 & 92--8500 & 50 & 48 \\
G24.60+0.08 & 3.45 & 193 & 41.3 & 0.52$\times$0.34 ($-61.8^\circ$) & 0.071$\times$0.045 ($39.4^\circ$) & 800 & 195 & 15.1--1400 & 92--8500 & 49 & 43 \\
G34.43+0.24 & 3.03 & 194 & 60.1 & 0.34$\times$0.32 ($-38.0^\circ$) & 0.072$\times$0.055 ($64.6^\circ$) & 1020 & 190 & 15.1--1300 & 92--8500 & 44 & 43 \\
G34.43+0.24MM2 & 3.03 & 147 & 34.5 & 0.34$\times$0.28 ($-79.0^\circ$) & 0.070$\times$0.054 ($-62.7^\circ$) & 1080 & 186 & 15.1--1400 & 92--8500 & 47 & 46 \\
G35.03+0.35A & 2.32 & 412 & 38.3 & 0.38$\times$0.25 ($-61.9^\circ$) & 0.077$\times$0.056 ($65.9^\circ$) & 1090 & 152 & 18.6--1100 & 92--8500 & 41 & 43 \\
G35.13-0.74 & 2.2 & 138 & 41.3 & 0.34$\times$0.27 ($-80.7^\circ$) & 0.069$\times$0.054 ($-61.5^\circ$) & 670 & 134 & 15.1--1400 & 92--8500 & 47 & 46 \\
\end{longtable}
\end{sidewaystable}

%% file: clump_information.tex
\begin{small}
\setlength{\tabcolsep}{4pt} 
\begin{longtable}{lcccccccccc}
    \caption{Properties of High-mass Cluster-forming Regions} \label{tab2} \\
    \toprule
    Clump name & R.A. (ICRS) & Decl. (ICRS) & Dist & T$_\mathrm{dust}$ & MF & CF & MF$_\mathrm{in~core}$ & MF$_\mathrm{orphan}$ & $\Sigma_\mathrm{cond}$ & R$_{eff}$\\
               & (hh:mm:ss.ss) & (dd:mm:ss.ss) & (kpc) & (K) & (\%) & (\%) & (\%) & (\%) & (pc$^{-2}$)& (au)\\
    (1) & (2) & (3) & (4) & (5) & (6) & (7) & (8) & (9) &(10) &(11)\\
    \midrule
    \endfirsthead

    \multicolumn{11}{l}{\tablename\ \thetable{} -- continued from previous page} \\
    \toprule
    Clump name & R.A. (ICRS) & Decl. (ICRS) & Dist & T$_\mathrm{dust}$ & MF & CF & MF$_\mathrm{in~core}$ & MF$_\mathrm{orphan}$ & $\Sigma_\mathrm{cond}$& R$_{eff}$\\
               & (hh:mm:ss.ss) & (dd:mm:ss.ss) & (kpc) & (K) & (\%) & (\%) & (\%) & (\%) & (pc$^{-2}$) & (au)\\
    (1) & (2) & (3) & (4) & (5) & (6) & (7) & (8) & (9) &(10) &(11)\\

    \midrule
    \endhead

    \midrule
    \multicolumn{11}{r}{\textit{Continued on next page}} \\
    \endfoot

    \bottomrule
    \endlastfoot
        G11.1-0.12 &  18:10:28.27  &  $-$19:22:30.92  & 3.00 & 15.8 & 33 $\pm$ 24 & 33 $\pm$ 24 & ... & 33 $\pm$ 24 & 40 $\pm$ 20 & 1140\\
        G11.92-0.61 &  18:13:58.02  &  $-$18:54:19.02  & 3.37 & 22.8 & 9 $\pm$ 8 & 9 $\pm$ 8 & ... & 17 $\pm$ 15 & 181 $\pm$ 43 & 1570\\
        G14.22-0.50S &  18:18:13.00  &  $-$16:57:21.82  & 1.90 & 15.8 & 19 $\pm$ 10 & 19 $\pm$ 10 & 25 $\pm$ 15 & 13 $\pm$ 12 & 201 $\pm$ 45 & 1520\\
        G24.60+0.08 &  18:35:40.50  &  $-$07:18:34.02  & 3.45 & 15.9 & 11 $\pm$ 10 & 11 $\pm$ 10 & 17 $\pm$ 15 & ... & 121 $\pm$ 34 & 2020\\
        G333.12-0.56 &  16:21:36.00  &  $-$50:40:50.01  & 3.30 & 16.1 & 50 $\pm$ 12 & 61 $\pm$ 11 & 57 $\pm$ 13 & 25 $\pm$ 20 & 332 $\pm$ 58 & 2310\\
        G333.23-0.06 &  16:19:51.20  &  $-$50:15:13.00  & 5.20 & 20.9 & 18 $\pm$ 8 & 23 $\pm$ 9 & 18 $\pm$ 11 & 18 $\pm$ 11 & 413 $\pm$ 64 & 2000\\
        G333.46-0.16 &  16:21:20.26  &  $-$50:09:46.56  & 2.90 & 24.6 & 53 $\pm$ 11 & 74 $\pm$ 10 & 67 $\pm$ 13 & 29 $\pm$ 16 & 252 $\pm$ 50 & 2390\\
        G335.579-0.272 &  16:30:58.76  &  $-$48:43:54.01  & 3.25 & 23.1 & 22 $\pm$ 13 & 22 $\pm$ 13 & 100 $\pm$ 25 & 13 $\pm$ 12 & 191 $\pm$ 44 & 1230\\
\end{longtable}
\end{small}

%% file: low_mass_cores_table.tex
\begin{longtable}{lcccccccc}
    \caption{Properties of Low Mass Cores}\\
    \label{tab3} \\
    \toprule
    Clump Name &Core Name & R.A. (ICRS) & Decl. (ICRS) & Flux & Number Density & Radius  & Category \\
              && (hh:mm:ss.ss) & (dd:mm:ss.ss) & (mJy) & (cm$^{-3}$)& (au) &  \\
     (1) &(2) & (3) & (4) & (5) & (6) & (7) & (8)  \\
    \midrule
    \endfirsthead

    \toprule
    Clump Name &Core Name & R.A. (ICRS) & Decl. (ICRS) & Flux & Number Density & Radius & Category \\
              && (hh:mm:ss.ss) & (dd:mm:ss.ss) & (mJy) & (cm$^{-3}$) & (au)& (au) &  \\
     (1) &(2) & (3) & (4) & (5) & (6) & (7) & (8) \\
    \midrule
    \endhead

    \bottomrule
    \endfoot
  G11.1-0.12 &ALMA2 & 18:10:28.30 & -19:22:30.60 & 3.3 & $2.0 \times 10^{9}$ & 340 & non-condensation \\
  G11.1-0.12 &ALMA3 & 18:10:28.10 & -19:22:35.27 & 2.7 & $1.4 \times 10^{8}$ & 770 & non-condensation \\
  G11.1-0.12 &ALMA4 & 18:10:28.05 & -19:22:35.70 & 0.9 & $4.0 \times 10^{8}$ & 380 & non-condensation \\
  G11.1-0.12 &ALMA5 & 18:10:28.20 & -19:22:33.75 & 1.9 & $1.3 \times 10^{8}$ & 710 & non-condensation \\
  G11.92-0.61 &ALMA3 & 18:13:58.14 & -18:54:16.24 & 7.2 & $4.7 \times 10^{8}$ & 660 & single \\
  G11.92-0.61 &ALMA4 & 18:13:58.98 & -18:54:18.40 & 3.6 & $9.0 \times 10^{8}$ & 420 & single \\
  G11.92-0.61 &ALMA5 & 18:13:58.06 & -18:54:14.49 & 9.3 & $3.2 \times 10^{8}$ & 820 & single \\
  G11.92-0.61 &ALMA6 & 18:13:57.93 & -18:54:08.91 & 5.7 & $2.9 \times 10^{8}$ & 720 & non-condensation \\
\end{longtable}

%% file: pybdsf-parameter.tex
\begin{table}[htb!]
  \centering
  \caption{Parameters for PyBDSF Package}
  \label{tabp}
  \begin{tabular}{@{}ll@{}}
    \toprule
    Parameter & Value \\
    \midrule
    adaptive\_rms\_box & True \\
    adaptive\_thresh & 50 \\
    rms\_box\_bright & (n\_pix\_beam, 0.5$\ast$n\_pix\_beam) \\
    mean\_map & \texttt{zero} \\
    rms\_box & (5$\ast$n\_pix\_beam, 0.5$\ast$n\_pix\_beam) \\
    thresh & \texttt{hard} \\
    thresh\_isl & 4.0 \\
    thresh\_pix & 5.0 \\
    \bottomrule
    
    \footnote{\texttt{adaptive\_rms\_box}: use when determining the RMS and mean maps.
 \texttt{adaptive\_thresh = 50}: represents the detection threshold of 50~$\sigma_{c8}$.
\texttt{rms\_box\_bright}: specifies the box size and step size used for RMS/mean map calculation near bright sources.
\texttt{rms\_box}: specifies the box size and step size used for RMS/mean map calculation. 
\texttt{n\_pix\_beam}: minimum number of pixels per beam.
\texttt{mean\_map}: means background mean map.
\texttt{thresh = `hard'}: means that sigma clipping is applied.
\texttt{thresh\_isl}: threshold of the island boundary, in units of sigma above the mean, used for fitting.
\texttt{thresh\_pix}: source detection threshold, and the threshold of the island peak in units of sigma above the mean.}
  \end{tabular}
  \vspace{0.5em}

\end{table}


  
  



%% file: low_mass_ysos.tex
\begin{longtable}{lccccc}
    \caption{Properties of Low-mass Condensation} \\
    \label{tab4} \\
    \toprule
    Clump Name & Condensation Name & R.A. (ICRS) & Decl. (ICRS) & Flux & Radius\\
              & & (hh:mm:ss.ss) & (dd:mm:ss.ss) & (10$^{-1}$mJy) & (au) \\  
     (1) &(2) & (3) & (4) & (5) &(6)  \\
    \midrule
    \endfirsthead

    \toprule
    Clump Name & Condensation Name & R.A. (ICRS) & Decl. (ICRS) & Flux & Radius\\
              & & (hh:mm:ss.ss) & (dd:mm:ss.ss) & (10$^{-1}$ mJy) & (au) \\  
     (1) &(2) & (3) & (4) & (5) &(6)  \\
    \midrule
    \endhead

    \bottomrule
    \endfoot
    \endlastfoot
  G11.1-0.12 &  ALMAe3 & 18:10:28.28 & -19:22:30.88 & 11.1 & 55 \\
  G11.1-0.12 &  ALMAe4 & 18:10:28.33 & -19:22:30.61 & 5.7 & 37 \\
  G11.1-0.12 &  ALMAe5 & 18:10:28.39 & -19:22:42.81 & 3.4 & 37 \\
  G11.1-0.12 &  ALMAe6 & 18:10:28.27 & -19:22:31.02 & 2.9 & 40 \\
  G11.92-0.61 &  ALMAe4 & 18:13:58.13 & -18:54:16.27 & 28.9 & 52 \\
  G11.92-0.61 &  ALMAe5 & 18:13:58.98 & -18:54:18.41 & 20.0 & 40 \\
  G11.92-0.61 &  ALMAe7 & 18:13:58.35 & -18:54:17.34 & 13.3 & 40 \\
  G11.92-0.61 &  ALMAe8 & 18:13:57.32 & -18:54:31.04 & 29.5 & 83 \\
\end{longtable}

%% file: low_mass_condensation_pairs.tex
\begin{longtable}{@{}llllllll@{}}\label{tabpp}\\
\caption{DIHCA low-mass condensation pairings} \\
\hline
Condensation pair & Separation & Separation (au) & $P_{\mathrm{cd}}$ & Clump name & Core name \\

&  (arcsec) &  (au) & (\%) &  &  \\
\hline
\endfirsthead

\hline
Condensation pair & Separation & Separation (au) & $P_{\mathrm{cd}}$ & Clump name & Core name \\

&  (arcsec) &  (au) & (\%) &  &  \\
\hline
\endhead
ALMAe3--ALMAe5 & 0.24 & 722.8 & 99.6\% & G11.1-0.12 & orphan \\
ALMAe13--ALMAe17 & 0.62 & 2105.7 & 59.5\% & G11.92-0.61 & orphan \\
ALMAe5--ALMAe19 & 0.1 & 191.2 & 99.7\% & G14.22-0.50S & ALMA7 \\
ALMAe8--ALMAe11 & 0.71 & 1355.8 & 88.2\% & G14.22-0.50S & ALMA5 \\
ALMAe13--ALMAe17 & 1.2 & 2282.2 & 71.8\% & G14.22-0.50S & orphan \\
ALMAe1--ALMAe7 & 0.33 & 1149.1 & 89.8\% & G24.60+0.08 & ALMA2 \\
ALMAe19--ALMAe23 & 0.08 & 231.3 & 98.7\% & G34.43+0.24 & orphan \\
ALMAe6--ALMAe12 & 0.08 & 242.0 & 98.5\% & G34.43+0.24 & ALMA6 \\
\end{longtable}